\documentclass[conference]{IEEEtran}
\usepackage{graphicx,amsmath,amssymb,color}
\usepackage{verbatim}
\usepackage[noadjust]{cite}
\usepackage{array}
\usepackage{cases}
\usepackage{url}
\usepackage{multirow}
\newtheorem{theorem}{Theorem}[section]
\newtheorem{proposition}[theorem]{Proposition}
\newtheorem{corollary}[theorem]{Corollary}
\newtheorem{lemma}[theorem]{Lemma}

\begin{document}
%
\title{Probabilistic bounds on the trapping redundancy of linear codes}

\author{\IEEEauthorblockN{Yu Tsunoda}
\IEEEauthorblockA{Department of Informatics and Imaging Systems\\Faculty of Engineering, Chiba University\\1-33 Yayoi-Cho Inage-Ku, Chiba 263-8522, Japan\\
Email: yu.tsunoda@chiba-u.jp}
\and
\IEEEauthorblockN{Yuichiro Fujiwara}
\IEEEauthorblockA{Graduate School of Advanced Integration Science\\Chiba University\\1-33 Yayoi-Cho Inage-Ku, Chiba 263-8522, Japan\\
Email: yuichiro.fujiwara@chiba-u.jp}}
\maketitle

\begin{abstract}
The trapping redundancy of a linear code is the number of rows of a smallest parity-check matrix such that no submatrix forms an $(a,b)$-trapping set.
This concept was first introduced in the context of low-density parity-check (LDPC) codes
in an attempt to estimate the number of redundant rows in a parity-check matrix suitable for iterative decoding.
Essentially the same concepts appear in other contexts as well such as robust syndrome extraction for quantum error correction.
Among the known upper bounds on the trapping redundancy,
the strongest one was proposed by employing a powerful tool in probabilistic combinatorics, called the Lov\'{a}sz Local Lemma.
Unfortunately, the proposed proof invoked this tool in a situation where an assumption made in the lemma does not necessarily hold.
Hence, although we do not doubt that nonetheless the proposed bound actually holds, for it to be a mathematical theorem,
a more rigorous proof is desired.
Another disadvantage of the proposed bound is that it is only applicable to $(a,b)$-trapping sets with rather small $a$.
Here, we give a more general and sharper upper bound on trapping redundancy
by making mathematically more rigorous use of probabilistic combinatorics without relying on the lemma.
Our bound is applicable to all potentially avoidable $(a,b)$-trapping sets with $a$ smaller than the minimum distance of a given linear code,
while being generally much sharper than the bound through the Lov\'{a}sz Local Lemma.
In fact, our upper bound is sharp enough to exactly determine the trapping redundancy for many cases,
thereby providing precise knowledge in the form of a more general bound with mathematical rigor.
\end{abstract}

\IEEEpeerreviewmaketitle

\section{Introduction}\label{intro}
Binary linear codes are the most extensively investigated class of error-correcting codes with many applications.
Mathematically speaking, an $[n,k,d]$ \textit{linear code} $\mathcal{C}$ of \textit{length} $n$, \textit{dimension} $k$, and \textit{minimum distance} $d$
is simply a $k$-dimensional subspace
of the $n$-dimensional vector space $\mathbb{F}_2^n$ over the finite field $\mathbb{F}_2$ of order $2$ such that
every nonzero vector in $\mathcal{C}$ is of weight at least $d$.
Thus, the study of $[n,k,d]$ linear codes may essentially be seen as a particular theory of vector spaces.

An interesting twist in coding theory is that the structures of the duals of subspaces are equally or sometimes even more important.
Indeed, it is often the case in modern coding theory that it is more important to find suitable sets of vectors in the duals of linear codes
than good linear codes themselves.
In the language of coding theory, we are often more interested in the properties of a \textit{parity-check matrix} $H$ of a linear code $\mathcal{C}$,
where $\mathcal{C} = \{\boldsymbol{c} \in \mathbb{F}_q^n \mid H\boldsymbol{c}^T = \boldsymbol{0}\}$, than in those of $\mathcal{C}$ itself.

A quintessential example in which parity-check matrices play the central role is \textit{low-density parity-check} (\textit{LDPC}) codes \cite{Richardson:2008}.
It is known that the combination of the efficient decoding scheme for LDPC codes, called \textit{belief propagation} (\textit{BP}) decoding,
and a compatible parity-check matrix makes remarkably low decoding complexity and high error correction capabilities simultaneously possible.

One natural direction of theoretical research on desirable parity-check matrices is
to focus on the kind of substructure that causes a given decoding method to fail.
For instance, given an $m \times n$ parity-check matrix $H$ of an $[n,k,d]$ linear code $\mathcal{C}$,
the main culprit of decoding errors in the case of basic syndrome decoding over the binary symmetric channel
is $m \times a$ submatrices for small $a$ in which every row is of even weight.
While this is simply a reworded version of the basic observation that
codewords $\boldsymbol{c} \in \mathcal{C}$ of low weight tend to cause decoding errors,
an interesting phenomenon in modern coding theory is that efficient, sophisticated decoding methods are
often susceptible to small particular submatrices that may not necessarily correspond to codewords.
Such substructures have been studied for various channels and decoding strategies under different names,
such as trapping sets \cite{Richardson:2003,Nguyen:2012,Karimi:2012}, near-codewords \cite{MacKay:2003b}, and stopping sets \cite{Richardson:2001d,Orlitsky:2005,Han:2008}.
This line of research has also made a bridge to a branch of combinatorial design theory with an extremal set theoretic flavor,
where particular substructures in a binary matrix have long been investigated as purely mathematical objects \cite{Colbourn:2009,Laendner:2010}.
For recent results in the context of LDPC codes, we refer the reader to \cite{Butler:2014} and references therein.

This paper studies the theoretical limit on the size of a parity-check matrix for a given linear code in which no small submatrices form
a special type of trapping set.
An $(a,b)$-\textit{trapping set} in an $m \times n$ binary matrix $H$ over $\mathbb{F}_2$ for $1 \leq a \leq n$ and $0 \leq b \leq m$ is an $m \times a$ submatrix $T$ of $H$
such that the number of rows of odd weight is exactly $b$.
An $(a,0)$-trapping set leads to a codeword of weight $a$.
In general, an $(a,b)$-trapping set corresponds to an error vector over a binary-in binary-out channel whose syndrome is of weight $b$.

A notable fact is that an $(a,b)$-trapping set for small $a$ can be removed rather easily if the size of a parity-check matrix is of no concern.
Indeed, adding a linearly dependent row to a parity-check matrix increases or at least maintains the weight of the syndrome of a binary error vector,
while keeping the subspace spanned by the rows unchanged.
For example, it is straightforward to see that a parity-check matrix that consists of all $2^{n-k}$ codewords of the dual of an $[n,k,d]$ linear code has no $(a,b)$-trapping set for all $1 \leq a \leq d-1$ and $0 \leq b \leq 2^{n-k-1}$ \cite{Laendner:2006}.
However, too large a parity-check matrix is problematic for various reasons,
such as increased decoding overhead in the case of LDPC codes \cite{MacKay:2003} and increased syndrome extraction failure rates for quantum error correction \cite{Ashikhmin:2014}.
Hence, given a linear code, we are interested in the smallest possible parity-check matrices that contain no undesirable $(a,b)$-trapping sets for all small $a$ and $b$ less than some given constants.

This limit was first investigated in \cite{Laendner:2009} as the \textit{trapping redundancy} of a linear code.
Their motivation was to estimate the number of rows of a smallest parity-check matrix
that is suited for BP decoding over the additive white Gaussian noise (AWGN) channel
in order to investigate how large a parity-check matrix for a good LDPC code should be.
They also proposed a very tight upper bound, which, as far as the authors are aware, has not yet been surpassed by any known general bound.

An important fact is that the proposed proof of the tightest bound relies on a very powerful tool in probabilistic combinatorics, called the \textit{Lov\'{a}sz Local Lemma}.
Unfortunately, as we will see later, the presented proof in its current form invokes the lemma
when an assumption required to apply the probabilistic tool does not necessarily hold,
although we do not doubt that the proposed bound holds for most linear codes of interest regardless of this subtle mathematical gap.

Aside from mathematical rigor, the proposed bound has a disadvantage that it is only applicable to $(a,b)$-trapping sets for relatively small $a$.
Any parity-check matrix for an $[n,k,d]$ linear code necessarily contains a nonzero codeword of the smallest weight as a $(d,0)$-trapping set,
while $(a,b)$-trapping sets with smaller $a$ and positive $b$ may not appear in a well-chosen parity-check matrix.
Thus, a most general bound would consider avoidance of $(a,b)$-trapping sets for all $a \leq d-1$.
However, the proposed bound only considers the case $a \leq \lfloor\frac{d-1}{2}\rfloor$.

The purpose of this paper is to give a sharp upper bound on the trapping redundancy of an $[n,k,d]$ linear code
that is mathematically rigorous and handles all $(a,b)$-trapping sets for $a \leq d-1$.
Our proof uses the same probability space as in \cite{Laendner:2009}.
However, our argument does not require the Lov\'{a}sz Local Lemma.
Instead, we make rigorous use of basic tools in probabilistic combinatorics to prove a much tighter and more general bound.
In fact, our bound shows that the trapping redundancy of an $[n,k,d]$ linear code often matches the trivial lower bound $n-k$ for many cases
in which all previously proved or proposed upper bounds are far from $n-k$.

It should be noted, however, that our results do not immediately give practical linear codes
because we do not take into account any other restrictions on a parity-check matrix that may arise in a real-life application.
Rather, our tight upper bound simply suggests that it is often not as difficult as previously thought
to avoid trapping sets in a parity-check matrix with few redundant rows if no other constraints are imposed.

In the next section, we define necessary notions and give a brief review on the known relevant results.
Our bound on trapping redundancy is proved in Section \ref{sec:main} as our main result.
Section \ref{sec:conclusion} concludes this paper with some remarks.

\section{Preliminaries}\label{pre}
Extensive empirical and theoretical research has shown that $(a,b)$-trapping sets for small $a$ and $b$ can greatly deteriorate
the performance of BP decoding over the AWGN channel,
making parity-check matrices with no small trapping sets more appealing (see \cite{Laendner:2009} and references therein).
The problem of avoiding small trapping sets also appears in robust quantum error correction under the phenomenological error model \cite{Fujiwara:2014a,Fujiwara:2015}
and erasure resilient coding for lage disk arrays \cite{Chee:2000,Muller:2004b}.
The common theme in these applications is that it is desirable for
a parity-check matrix to have as few redundant rows as possible but contain no small trapping sets.

The $(a,b)$-\textit{trapping redundancy} $T_{a,b}(\mathcal{C})$ of an $[n,k,d]$ linear code $\mathcal{C}$
is the number of rows of a smallest parity-check matrix for $\mathcal{C}$ that contains no $(a,t)$-trapping set for $0 \leq t \leq b-1$.
The \textit{collective} $(a,b)$-trapping redundancy $\overline{T}_{a,b}(\mathcal{C})$ of $\mathcal{C}$
is the number of rows of a smallest parity-check matrix for $\mathcal{C}$ that contains no $(s,t)$-trapping set for
$1 \leq s \leq a$ and $0 \leq t \leq b-1$.
Trivially, $T_{a,b}(\mathcal{C}) \leq \overline{T}_{a,b}(\mathcal{C})$ for any linear code $\mathcal{C}$ and any parameters $a$ and $b$.

We use a special kind of combinatorial matrix to employ probabilistic proving methods.
An \textit{orthogonal array} $\textup{OA}(m,n,l,s)$ is an $m\times n$ matrix over a finite set $\Gamma$ of cardinality $l$ such that
in any $m\times s$ submatrix every $s$-dimensional vector in $\Gamma^s$ appears exactly $\frac{m}{l^s}$ times as a row.
This definition demands that $m$ be divisible by $l^s$.
A simple but useful observation is that an $\textup{OA}(m,n,l,s)$ for $s\geq2$ is also an $\textup{OA}(m,n,l,s-i)$ for any $0 \leq i \leq s-1$.

The following is an immediate corollary of Delsarte's equivalence theorem \cite[Theorem 4.5]{Delsarte:1973}.
\begin{proposition}\label{prop:linearDelsarte}
Let $\mathcal{C}$ be a linear code of length $n$, dimension $k$, and minimum distance $d$.
A $2^{n-k}\times n$ matrix formed by all codewords of $C^\perp$ as rows is an $\textup{OA}(2^{n-k},n,2,d-1)$.
\end{proposition}

Informally, the above proposition shows that for $1 \leq a \leq d-1$, fixed $a$ bits in a randomly chosen codeword in $\mathcal{C}^\perp$ look completely random.
Through this observation, it is claimed in \cite{Laendner:2009} that for an $[n,k,d]$ linear code $\mathcal{C}$ and positive integer $a \leq \lfloor\frac{d-1}{2}\rfloor$,
the $(a,b)$-trapping redundancy $T_{a,b}(\mathcal{C})$ would be smaller than or equal to $m+n-k-1$,
where $m$ is the smallest integer such that
\[2^{-m}\left(\binom{n}{a}-\binom{n-a}{a}\right)\sum_{j=0}^{b-1}\binom{m}{j} \leq \frac{1}{e}\]
with $e$ being the base of the natural logarithm.
While this inequality does not consider the case $\lfloor\frac{d-1}{2}\rfloor < a \leq d-1$,
it is by far the tightest for the case when $1 \leq a \leq \lfloor\frac{d-1}{2}\rfloor$.

Unfortunately, the proposed proof relies on the following well-known lemma in a way
a more mathematically rigorous argument is desirable for the bound to be considered a mathematical theorem.
\begin{proposition}[Lov\'{a}sz Local Lemma]\label{LLM}
Take a finite set of events $A_i$ in an arbitrary probability space
such that each $A_i$ occurs with probability at most $p$ and is mutually independent of all others except for at most $x$ of the others.
If $(x+1)ep\leq1$, the probability that none of $A_i$ occurs is positive.
\end{proposition}

The proposed proof starts with taking $m$ codewords independently and uniformly at random from the dual $\mathcal{C}^\perp$
to form an $m \times n$ potential parity-check matrix $H$ and then invokes the Lov\'{a}sz Local Lemma given above
to assert that there is a positive probability that no $m \times a$ submatrix forms an $(a,t)$-trapping set for $0 \leq t \leq b-1$.
The final step is to take $n-k-1$ more rows from $\mathcal{C}^\perp$
to make sure that the rank of the resulting $(m-n-k-1) \times n$ matrix is $n-k$.

To see a subtle gap in the above argument, for $1 \leq a\leq \lfloor\frac{d-1}{2}\rfloor$
let $\mathcal{M}$ be the set of $m \times a$ submatrices in $H$
and $\mathcal{N} \subset \mathcal{M}$ its subset such that any pair $N, N' \in \mathcal{N}$ of submatrices in $\mathcal{N}$ share no columns.
Define $A_M$ to be the event that $m \times a$ submatrix $M \in \mathcal{M}$ forms an $(a,t)$-trapping set.
The problematic part of the proposed proof is that it invokes the Lov\'{a}sz Local Lemma by assuming that the events $A_N$ for $N \in \mathcal{N}$ are always mutually independent, which is, strictly speaking, not true.
Indeed, while Proposition \ref{prop:linearDelsarte} assures that $A_N$ for $N \in \mathcal{N}$ are pairwise independent,
this fact does not imply that they are mutually independent in general.
Although a variant of the Lov\'{a}sz Local Lemma which does not assume mutual independence is also known in probabilistic combinatorics (see \cite{Alon:2008}),
as is also pointed out in \cite{Han:2008},
it seems unlikely for the Lov\'{a}sz Local Lemma and its variants to be able to give such a strong bound in this probability space.
Nonetheless, in the next section we show that basic tools in probabilistic combinatorics can prove an even stronger and more general bound.

\section{Bounds by probabilistic combinatorics}\label{sec:main}
Now we present an upper bound on trapping redundancy without relying on the Lov\'{a}sz Local Lemma.
In what follows, for a pair $x, y$ of nonnegative integers $x \geq y$,
\[
{x \brack y}_q = \prod_{i=0}^{y-1}\frac{1-q^{x-i}}{1-q^{i+1}}
\]
is defined to be the Gaussian binomial coefficient.

Our argument provides an explicit upper bound on the collective $(a,b)$-trapping redundancy $\overline{T}_{a,b}(\mathcal{C})$
for an arbitrary $[n,k,d]$ linear code $\mathcal{C}$.
\begin{theorem}\label{MainTh1}
Let $\mathcal{C}$ be an $[n, k, d]$ linear code.
For $1 \le a \le d-1$ and $b \geq 0$,
\begin{align*}
\overline{T}&_{a,b}(\mathcal{C}) \leq \min_{t \in \mathbb{N}}\left\{t + \left\lfloor2^{-t}\sum_{u=1}^{a}\binom{n}{u}\sum_{i=0}^{b}i\binom{t}{b-i}
\right.\right.\\
&+\left.\left.2^{-t(n-k)}\sum_{r=0}^{n-k}(n-k-r){n-k \brack r}_2\prod_{i=0}^{r-1}(2^t-2^i)\right\rfloor\right\}.
\end{align*}
\end{theorem}

To prove the above theorem, we employ the following well-known fact.
\begin{lemma}\label{LemmaForMainTh1}
Let $\mathcal{C}$ be an $[n, k, d]$ binary linear code and $H_t$ a $t \times n$ matrix of which each row is drawn independently and uniformly at random from the dual $\mathcal{C}^{\perp}$. For $1 \le r \le n-k$, the probability that $H_t$ is of rank $r$ is 
\[
\frac{{n-k \brack r}_2\prod_{i=0}^{r-1}(2^t-2^i)}{2^{t(n-k)}}.
\]
\end{lemma}
For various known proofs of the above lemma, see, for example, \cite{Fisher:1966,Kovalenko:1967}.

We now prove Theorem \ref{MainTh1}.
In what follows, the expected value of a given random variable $X$ is denoted by $\mathbb{E}(X)$.
\begin{IEEEproof}[Proof of Theorem \ref{MainTh1}]
Let $H_t$ be a $t \times n$ matrix whose rows are drawn from $\mathcal{C}^{\perp}$ independently and uniformly at random.
Define $\mathcal{M}_{t,u}$ to be the set of $t \times u$ submatrices in $H_t$.
Note that Proposition \ref{prop:linearDelsarte} implies that for any $1\times u$ submatrix $\boldsymbol{u}$ in $H_t$ with $u \leq d-1$,
the probability that $\boldsymbol{u}$ is of odd weight is $\frac{1}{2}$.
For $M \in \mathcal{M}_{t,u}$, let $w_M$ be the random variable counting the number of rows of odd weight in $M$.
Define $X_M$ to be the random variable
\[X_M = \begin{cases}
0 &\text{if } w_M \geq b,\\
b-w_M &\text{otherwise.}
\end{cases}
\]
Note that $X_M$ counts the smallest number of additional rows required to turn $M$ into a $(u,c)$-trapping set with $c\geq b$.
Let
\[Y_t = n-k-\operatorname{rank}(H_t)\]
be the random variable counting the smallest number of additional rows required to turn $H_t$ into a parity-check matrix for $\mathcal{C}$.
Define
\[Z_t = Y_t+\sum_{u=1}^a\sum_{M\in\mathcal{M}_{t,u}}X_M.\]
Note that we can construct a parity-check matrix for $\mathcal{C}$ which contains no $(u,v)$-trapping set for all $1 \leq u \leq a$ and $0 \leq v \leq b-1$
by adding to $H_t$ at most $Z_t$ codewords of $\mathcal{C}^\perp$ as rows,
which means that there exists a $(t+\lfloor\mathbb{E}(Z_t)\rfloor)\times n$ parity-check matrix for $\mathcal{C}$
which contains no $(u,v)$-trapping set for all $1 \leq u \leq a$ and $0 \leq v \leq b-1$.
Hence, we have
\begin{align}\label{ineqT}
\overline{T}_{a,b}(\mathcal{C})\leq \min_{t\in\mathbb{N}}\left\{t+\lfloor\mathbb{E}(Z_t)\rfloor\right\}.
\end{align}
To calculate the expected value on the right-hand side, notice that for $M \in \mathcal{M}_{t,u}$,
\begin{align*}
\mathbb{E}(X_M) &= \sum_{i=1}^{b}i2^{-(b-i)}2^{(b-i)-t}\binom{t}{b-i}\\
&= 2^{-t}\sum_{i=1}^{b}i\binom{t}{b-i}.
\end{align*}
By Lemma \ref{LemmaForMainTh1},
\[\mathbb{E}(Y_t) = 2^{-t(n-k)}\sum_{r=0}^{n-k}(n-k-r){n-k \brack r}_2\prod_{i=0}^{r-1}(2^t-2^i).\]
Thus, by linearity of expectation, we have
\begin{align*}
\mathbb{E}(Z_t) &= \mathbb{E}(Y_t)+\sum_{u=1}^{a}\sum_{M\in\mathcal{M}_{t,u}}\mathbb{E}(X_M)\\
&= 2^{-t}\sum_{u=1}^{a}\binom{n}{u}\sum_{i=0}^{b}i\binom{t}{b-i}\\
&+2^{-t(n-k)}\sum_{r=0}^{n-k}(n-k-r){n-k \brack r}_2\prod_{i=0}^{r-1}(2^t-2^i).
\end{align*}
Plugging in the above equation into (\ref{ineqT}) proves the assertion.
\end{IEEEproof}

The upper bound we just proved is extremely tight for quite a large portion of known linear codes.
In fact, Theorem \ref{MainTh1} shows that the trivial lower bound $\overline{T}_{a,b}(\mathcal{C}) \geq n-k$ is
indeed the true collective trapping redundancy for many $[n,k,d]$ linear codes $\mathcal{C}$.
The following immediate corollary is useful for checking whether the collective trapping redundancy of a given linear code matches the trivial lower bound.
\begin{corollary}\label{CTisn-k}
Let $\mathcal{C}$ be an $[n,k,d]$ linear code. If 
\begin{align*}
&2^{-(n-k)}\sum_{u=1}^{a}\binom{n}{u}\sum_{i=0}^{b}i\binom{n-k}{b-i}\\
&+2^{-(n-k)^2}\sum_{r=0}^{n-k}(n-k-r){n-k \brack r}_2\prod_{i=0}^{r-1}(2^{n-k}-2^i) < 1,
\end{align*}
then $\overline{T}_{a,b}(\mathcal{C}) = n-k$.
\end{corollary}
\begin{IEEEproof}
Take exactly $n-k$ rows independently and uniformly at random from $\mathcal{C}^\perp$
and follow the same argument as in the proof of Theorem \ref{MainTh1}.
\end{IEEEproof}

Because the left-hand side of the inequality in the above corollary is exponentially small for fixed $a$ and $b$,
if we fix the rate $\frac{k}{n}$, taking a longer linear code ensures that a parity-check matrix with no redundant row can avoid all $(a,b)$-trapping sets.
To see the usefulness of Corollary \ref{CTisn-k}, recall the following basic formulation of the Gilbert-Varshamov bound
(see \cite{Gaborit:2008} for recent progress on bounds of Gilbert-Varshamov type).
\begin{theorem}[Gilbert-Varshamov bound]\label{th:GV}
An $[n,k,d]$ linear code exists if
\[2^{n-k}\geq\sum_{i=0}^{d-1}\binom{n}{i}.\]
\end{theorem}

While this existence result has been known for more than sixty years,
it is quite difficult to beat and still serves as a quick benchmark for goodness of a code today.
Now, if an $[n,k,d]$ linear code $\mathcal{C}$ obeys the Gilbert-Varshamov bound,
the first term $2^{-(n-k)}\sum_{u=1}^{a}\binom{n}{u}\sum_{i=0}^{b}i\binom{n-k}{b-i}$ on the left-hand side of the inequality in Corollary \ref{CTisn-k} tends to $0$ exponentially fast as $n-k$ increases.
Since the second term is also exponentially small,
the upper bound on $\overline{T}_{a,b}(\mathcal{C})$ quickly becomes $n-k$ once the size of $\mathcal{C}$ goes below the Gilbert-Varshamov bound.
This simple observation also shows the existence of an asymptotically good sequence of linear codes with constant relative distance and the lowest possible collective trapping redundancy.

It is also notable that Theorem \ref{MainTh1} is a bound on the collective $(a,b)$-trapping redundancy $\overline{T}_{a,b}(\mathcal{C})$,
which implies that it also serves as an upper bound on the $(a,b)$-trapping redundancy $T_{a,b}(\mathcal{C})$
because $T_{a,b}(\mathcal{C}) \leq \overline{T}_{a,b}(\mathcal{C})$ by definition.
While known upper bounds on the trapping redundancy including the one relying on the Lov\'{a}sz Local Lemma can not match the trivial lower bound in general,
Theorem \ref{MainTh1} can verilify that $T_{a,b}(\mathcal{C}) = n-k$ by showing the much stronger statement that $\overline{T}_{a,b}(\mathcal{C}) = n-k$ for all sufficiently large linear codes $\mathcal{C}$.
For instance, an upper bound on the $(a,b)$-trapping redundancy of the Margulis code \cite{Margulis:1982} of length $2640$ and dimension $1320$ was derived in \cite{Laendner:2009} for $a \leq 14$ and $b = 5$ by using the Lov\'{a}sz Local Lemma as example cases.
Table \ref{Margulis} lists the upper bound by the lemma along with our upper bound by Theorem \ref{MainTh1} and the trivial lower bound.
\begin{table}
\renewcommand{\arraystretch}{1.3}
\caption{$(a,b)$-Trapping redundancy of the Margulis code}
\label{Margulis}
\centering
\begin{tabular}{ccccc}
\hline\hline
$a$ & $b$ & Trivial lower bound & Upper bound by Theorem \ref{MainTh1} & LLL\rlap{\textsuperscript{a}}\\
\hline
$6$ & $5$ & $1320$ & $1320$ & $1394$\\
$8$ & $5$ & $1320$ & $1320$ & $1413$\\
$12$ & $5$ & $1320$ & $1320$ & $1448$\\
$14$ & $5$ & $1320$ & $1320$ & $1464$\\
 \hline
 \hline
\multicolumn{5}{l}{\scriptsize\textsuperscript{a}
This column lists the upper bound in \cite{Laendner:2009} using the Lov\'{a}sz Local Lemma.}\\
\end{tabular}
\end{table}
Because the length and dimension of the code are $2640$ and $1320$,
the trapping redundancy must be at least $2640-1320=1320$.
As shown in the table, for all examined cases, Theorem \ref{MainTh1} determines the exact trapping redundancy
by showing that even the collective trapping redundancy is already $1320$.

\section{Concluding remarks}\label{sec:conclusion}
We have derived a tight upper bound on the collective trapping redundancy of a linear code by employing probabilistic combinatorics.
An immediate corollary showed that the collective $(a,b)$-trapping redundancy of an $[n,k,d]$ linear code whose dimension is strictly below the Gilbert-Varshamov bound matches the trivial lower bound $n-k$ unless $n$ is too small and $b$ is too large.
Our bound is applicable to all $1 \leq a \leq d-1$, which is exactly the range within which collective $(a,b)$-trapping redundancy is well-defined.

It should be noted, however, that our optimistic results do not necessarily imply that there exist practical linear codes for a specific real-life application where low trapping redundancy is desirable.
Indeed, for the theory of trapping redundancy to be practical, it is necessary to take into account other restrictions that may arise in practice.

Another related problem we did not address is efficient algorithms for constructing parity-check matrices whose existence is proved by our probabilistic argument.
Ideally, such derandomization should be carried out while ensuring that the resulting parity-check matrices with no small trapping sets possess other desirable properties as well.
For instance, in the context of LDPC codes, one key requirement is that a parity-check matrix has only a very small number of nonzero entries.
From this point of view, it may also be a very promising path to generalize the approach taken in \cite{Laendner:2009}.
Indeed, their current approach already gives a fairly sparse matrix because we only need to add a small number of extra rows to a sparse parity-check matrix.

As we have seen, the probabilistic proof presented here shows that a parity-check matrix with no redundant rows can often completely avoid small trapping sets.
However, our theoretical analysis does not immediately give a solution to a practical problem.
Therefore, our results are simply the first step that showed that trapping redundancy can be much smaller than previously thought.
Nevertheless, we hope that the results presented here stimulate research on trapping sets with applications in mind and related theoretical problems.


\end{document}